**Insulator-like behavior coexisting with metallic electronic structure in strained FeSe thin films grown by molecular beam epitaxy**

Kota Hanzawa,[1] Yuta Yamaguchi,[1] Yukiko Obata,[1,a)] Satoru Matsuishi,[2] Hidenori Hiramatsu,[1,2,b)] Toshio Kamiya,[1,2] and Hideo Hosono[1,2]

[1] *Laboratory for Materials and Structures, Institute of Innovative Research, Tokyo Institute of Technology, Mailbox R3-3, 4259 Nagatsuta-cho, Midori-ku, Yokohama 226-8503, Japan*

[2] *Materials Research Center for Element Strategy, Tokyo Institute of Technology, Mailbox SE-1, 4259 Nagatsuta-cho, Midori-ku, Yokohama 226-8503, Japan*

This paper reports that ~10-nm-thick iron selenide (FeSe) thin films exhibit insulator-like behavior in terms of the temperature dependence of their electrical resistivity even though bulk FeSe has a metallic electronic structure that has been confirmed by photoemission spectroscopy and first-principles calculations. This apparent contradiction is explained by potential barriers formed in the conduction band. Very thin FeSe epitaxial films with various atomic composition ratios ([Fe]/[Se]) were fabricated by molecular beam epitaxy and classified into two groups with respect to lattice strain and electrical properties. Lattice parameter *a* increased and lattice parameter *c* decreased with increasing [Fe]/[Se] up to 1.1 and then *a* levelled off and *c* began to decrease at higher [Fe]/[Se]. Consequently, the FeSe films had the most strained lattice when [Fe]/[Se] was 1.1, but these films had the best quality with respect to crystallinity and surface flatness. All the FeSe films with [Fe]/[Se] of 0.8–1.9 exhibited insulator-like behavior, but the temperature dependences of their electrical resistivities exhibited different activation energies $E_a$ between the Se-rich and Fe-rich regions; i.e., $E_a$ were small (a few meV) up to [Fe]/[Se]=1.1 but jumped up to ~25 meV at higher [Fe]/[Se]. The film with [Fe]/[Se]=1.1 had the smallest $E_a$ of 1.1 meV and exhibited an insulator–superconducting transition at 35 K with zero resistance under gate bias. The large $E_a$ of the Fe-rich films was attributed to the unusual lattice strain with tensile in-plane and relaxed out-of-plane strains. The large $E_a$ of films with [Fe]/[Se]>1.1 resulted in low mobility with a high potential barrier of ~50 meV in the conduction band for percolation carrier conduction compared with that of the [Fe]/[Se]=1.1 film (~17 meV). Therefore, the Fe-rich films exhibited remarkable insulator-like behavior similar to a semiconductor despite their metallic electronic structure. The high potential barrier of Fe-rich films is tentatively attributed to the presence of large amounts of excess Fe, which could plausibly cause a broad superconducting transition without zero resistance under gating.

___________________________

[a] Present address: Photon Factory and Condensed Matter Research Center, Institute of Materials Structure Science, High Energy Accelerator Research Organization (KEK), Tsukuba, Japan

[b] Author to whom correspondence should be addressed. Electronic mail: h-hirama@mces.titech.ac.jp

I. Introduction

The iron (Fe)-based superconductor LaFeAsO$_{1-x}$F$_x$ was discovered in early 2008 [1] and has invoked intensive effort to explore new related superconductors and investigate their physical properties and the origin/mechanism of their superconductivity [2−4]. The parent phases of almost all Fe-based superconductors exhibit "poor (bad)" metallic electronic conduction and have long-range antiferromagnetic ordering, which are considered to be closely related to the appearance of superconductivity; i.e., carrier doping of the parent phase suppresses and finally removes the magnetic ordering, leading to the appearance of superconductivity. The superconducting critical temperature ($T_c$) of this family has reached 55 K for a bulk state of SmFeAsO$_{1-x}$F$_x$ [5]. The conventional F$^-$ doping of the O$^{2-}$ site in the parent LaFeAsO (LaFeAsO$_{1-x}$F$_x$) exhibits only one superconducting dome in its electronic phase diagram, as seen for cuprate superconductors. In contrast, H$^-$ doping (LaFeAsO$_{1-x}$H$_x$) extends the doping limit and leads to the appearance of a second superconducting dome in the high-doping region, clarifying that more than one parent phase exists in the phase diagrams of Fe-based superconductors [6, 7]. Therefore, it is considered that the parent phases have essential roles in the superconducting mechanisms and determine $T_c$, similar to the case of high-$T_c$ cuprate superconductors, in which a Mott-insulator antiferromagnetic parent phase with strong electron correlation plays a critical role in their high-$T_c$ superconductivity [8].

Among Fe-based superconductors, tetragonal FeSe [9] is a unique compound from the viewpoints of both crystal and electronic structures. The crystal structure of FeSe consists of a simple stack of anti-PbO-type FeSe layers composed of edge-sharing $FeSe_4$ tetrahedra; i.e., unlike the other Fe-based superconductors with complex chemical compositions such as 1111-type LaFeAsO, in which insulating LaO layers are inserted into the conducting FeAs layers, FeSe has no insertion layer between the FeSe layers. Even though its $T_c$ is as low as 8 K in the bulk state, the electronic transport properties of FeSe, especially $T_c$, can be strongly modulated by various external parameters such as chemical composition, synthesis conditions, pressure, and substrate [10−12]. For instance, the chemical pressure (i.e., strain) induced by isovalent substitution of Te at the Se site and an external high pressure raised $T_c$ to 19 K for $FeSe_{0.5}Te_{0.5}$ [13] and 37 K for FeSe at 8.9 GPa [14], respectively. $T_c$ of thin films can be enhanced compared with that of the bulk state. The $T_c$ of an FeSe film deposited on a $CaF_2$ substrate was raised to 11.4 K [15] because the compressive in-plane lattice strain in the thin film had a similar effect to external pressure. Conversely, when FeSe films were deposited on $SrTiO_3$ (STO) and MgO substrates, the electronic transport properties of the films exhibited interesting thickness dependences. For thick films (≥200 nm), $T_c$ were almost the same as that of the bulk (~8 K). In contrast, for thin films (≤50 nm), superconductivity disappeared [16] and insulator-like behavior (i.e., increasing resistivity with decreasing temperature) was observed [17, 18]. Although the insulator-like behavior has been proposed to originate from the lattice strain in the films [16] and/or highly textured FeSe surfaces (i.e., the coexistence of nonsuperconducting and granular superconducting phases) [19, 20], the physical origin and mechanism of the insulator-like behavior of thin FeSe films are still under debate.

Starting from thin FeSe films with an insulator-like state, we successfully induced an insulator–superconductor phase transition by electrostatic electron doping using an electric double-layer transistor (EDLT) structure with an ionic liquid as a gate insulator [21, 22]. $T_c$ of the EDLT was raised to 35 K under an applied gate bias, which is ca. four times higher than that of bulk FeSe and agrees with similar recent reports [23−26]. In EDLTs with insulator-like FeSe layers, we proposed that the insulator-like FeSe can be an effective parent phase because its strong electron correlation should lead to higher $T_c$, like cuprate superconductors; this prediction was initially investigated using insulating $TlFe_{1.6}Se_2$ [27]. However, the superconducting properties of the insulator-like FeSe EDLTs strongly depended on their growth conditions and film structure; e.g., the films grown at rates that were too high or too low exhibited lower $T_c$ without zero resistance than that of films grown at the optimal rate. We tentatively concluded that over- or underdoped states originating from the defect structures formed under the sub-optimum growth conditions is one of the origins of the poor superconducting properties of the EDLTs [22]. However, more detailed examination is still required to clarify the underlying physical mechanism and improve the superconducting properties of FeSe thin films. In addition, it has been reported that an extremely high $T_c$ of ≥77 K was induced by optimum thermal annealing of the initial insulating state of very thin (one-unit-cell thick) FeSe films on STO substrates [28, 29]. Therefore, an insulator-like FeSe parent phase is essential to potentially realize high $T_c$. An approach using thin-film growth processes is the most effective way to examine such effects because the insulator-like behavior of FeSe is affected by several factors such as lattice strain, surface texture, and/or sample dimensionality (i.e., sample thickness).

Superconductivity in bulk FeSe has already been investigated in detail including the effects of chemical composition. It has been reported that excess Fe at the interstitial sites does not affect $T_c$ [30, 31] or suppresses $T_c$ through segregation of excess Fe [32] because the possible interstitial Fe concentration is limited to ~5%. In the Se-rich case, ordered or disordered Fe vacancies are formed depending on the fabrication conditions. If the ordered Fe vacancies are quenched, the primitive unit cell is extended to a √2×√2×1 superlattice structure [33]. The phase with the Fe-vacancy superstructure also exhibits insulator-like behavior. These findings suggest that if we can vary the chemical composition and superstructures of FeSe films over a much wider range, e.g., via nonequilibrium thin-film growth processes, it may be possible to discover novel phenomena such as the two superconducting domes in LaFeAs($O_{1-x}H_x$) [6, 7]. Moreover, the strain introduced during thin-film growth should raise the potential barrier for conduction carriers (i.e., insulating behavior) in insulator-like FeSe because of the intrinsic sensitivity of the physical properties of FeSe to its local structure. Thus, the investigation of the physical properties, particularly the electronic transport properties, of insulator-like FeSe over a wide range of chemical compositions is important to find a way to achieve much higher $T_c$ by carrier doping using an EDLT structure as well as atomically thin FeSe layers.

In this paper, we focus on insulator-like states of FeSe thin films with a thickness of ~10 nm and investigate the electronic transport properties of these films. It is found that there is a chemical composition boundary for conductivity activation energy, which also corresponds to strain in the films. The activation energies of the samples with excess Fe are approximately one order of magnitude higher than those of the films that are deficient in Fe. We clarify that the origin of this drastic change in the activation energies of the films is the different potential barrier heights in the conduction band for percolation conduction of carriers, and that the reasons for the different potential barriers are lattice variation and excess Fe in the FeSe lattice introduced during the nonequilibrium thin-film growth process.

## II. Experimental procedures

Epitaxial FeSe thin films were grown by molecular beam epitaxy (MBE) using an EV-100/PLD-S growth chamber (Eiko, Japan) under a base pressure of $\leq 1\times 10^{-7}$ Pa. An (00$l$)-oriented STO single crystal was used as a substrate, which was thermally annealed at 1050 °C in air after etching in buffered HF to obtain an atomically flat surface prior to film deposition [34]. The STO substrate was heated in the MBE growth chamber using an infrared semiconductor laser (LU0915C300-6, Lumics GmbH, Germany, $\lambda$=915 nm) through a thin backplate made of stainless steel to absorb the infrared laser irradiation and raise the substrate temperature ($T_s$) from 350 to 700°C, which was calibrated using a thermocouple directly connected to a sapphire single-crystal plate. Fe (99.99%) and Se (99.999%) were evaporated from separate Knudsen cells and the flux of each element was controlled by the cell temperature with a beam flux monitor positioned just beneath the substrate.

Crystalline phase and structure were characterized along the out-of-plane (i.e., vertical diffraction to the substrate surface) and in-plane (i.e., parallel to) directions by x-ray diffraction (XRD) using a SmartLab diffractometer (Rigaku, Japan) with Cu K$\alpha_1$ radiation that was monochromated with a two-bounce Ge (220) crystal. The geometry of these axes of the XRD apparatus can be found in Ref. [35]. *In situ* reflection high energy electron diffraction (RHEED) observation at an acceleration voltage of 20 kV was also performed in the MBE growth chamber to confirm the heteroepitaxial growth of FeSe. Film crystallinity was evaluated using the rocking-curve full width at half maximum (FWHM) values of the out-of-plane 001 ($\Delta\omega$) and in-plane 200 ($\Delta\phi$) diffractions. In the out-of-plane rocking-curve measurements, an additional two-bounce Ge (220) crystal was mounted in front of the scintillation detector to achieve the higher angle-dispersion resolution of $\Delta\omega$<0.001° instead of the usual optics with $\Delta\omega$<0.01°. The surface roughness was characterized by the root-mean-square roughness ($R_{rms}$) estimated from surface morphology in topography images captured by atomic force microscopy (AFM) on a MutiMode8 scanning probe microscope (Bruker Nano Inc., USA) using a tapping mode and optical cantilever. Film thickness was controlled in the ranges of 15–20 nm (to optimize $T_s$) and ~10 nm (to optimize chemical composition) to grow films that exhibited insulator-like behavior. We chose these thicknesses in this study because electrically continuous and uniform films were obtained for films thicker than 5.5 nm. Film thickness was precisely determined by x-ray reflectivity (XRR) or analysis of interference fringes around the FeSe 001 out-of-plane diffraction peak. For films with rough surfaces (i.e., thickness was not able to be determined by XRR and XRD), we used a stylus profiler (Alpha-Step D-120, KLA-Tencor, USA) to roughly estimate their thicknesses. Chemical compositions of the films (i.e., the atomic ratio of Fe to Se, [Fe]/[Se]) were determined by wavelength-dispersive x-ray fluorescence (XRF) using an S8 Tiger spectrometer (Bruker AXS, Germany). The XRF signals for Fe and Se were initially calibrated using quantitative analysis results of electron probe microanalysis obtained with a JXA-8530F analyzer (JEOL, Japan) because we confirmed the quantitative reliability of this technique in our previous work [21]. To determine the chemical compositions of large lateral-size segregated grains, field-emission scanning electron microscopy (FE-SEM) using a JSM-7600F electron microscope (JEOL) with an energy-dispersive x-ray (EDX) detector in point-analysis mode was employed. The spatial resolution was comparable to the incident beam size (<20 nm) because the films were so thin that extra fluorescence originating from secondary electrons scattered in the film did not occur.

Electronic transport properties of the obtained films were characterized by the temperature ($T$) dependence of their longitudinal resistivity ($\rho_{xx}=\rho$) and Hall-effect measurements (i.e., transverse resistivity, $\rho_{xy}$) with a physical property measurement system (Quantum Design Inc., USA) at 4.2–300 K. In the $\rho$–$T$ measurements, we used a four-probe geometry with Au electrodes deposited by direct-current sputtering using an SPF-332HS sputtering system (Canon Anelva, Japan). Hall-effect measurements were conducted using a six-terminal Hall bar structure (500 μm long and 200 μm wide). A shadow mask was used in the patterning and lift-off process to form electrodes. An external magnetic field of up to $\mu_0 H = \pm 3$ T with an interval of 0.5 T was applied. To cancel out the offset effects, we calculated the net $\rho_{xy}$ using the relation $\rho_{xy}=(\rho_{xy}^{+}-\rho_{xy}^{-})/2$, where $\rho_{xy}^{+}$ and $\rho_{xy}^{-}$ were measured under magnetic fields with opposite polarities.

Electronic structures were observed directly by angle-resolved photoemission spectroscopy (ARPES) excited by monochromatic He Iα light with a photon energy of 21.2180 eV from an MBS L-1 discharge lamp (MB Scientific AB, Sweden). Before the ARPES measurements, we constructed an all-*in situ* sample transfer system [36] (i.e., to prevent sample exposure to any gasses such as air, pure $N_2$, or Ar), in which the growth chamber, ARPES measurement chamber, and sample carrier transfer chambers were connected in an ultrahigh vacuum of $<1\times10^{-7}$ Pa because of the serious sensitivity of FeSe surfaces to air [37]. The ARPES measurements were carried out under an ultrahigh vacuum of $<4\times10^{-8}$ Pa and at a low temperature of ~10 K. We scanned along the Γ−M line using a Scienta DA30 photoelectron analyzer (Scienta Omicron Inc., Germany). The detector resolutions were set to 10 meV for energy and 1° for angle. To precisely evaluate binding energies from the obtained results, the Fermi level ($E_F$) was calibrated by measuring the Fermi edge of a polycrystalline Au.

III. Results and discussion

*III-1. Phase diagram of FeSe growth on STO (001)*

To easily understand the entire discussion, we first illustrate the relationships between the obtained crystalline phases, crystallite orientations, [Fe]/[Se], and $T_s$ in Fig. 1(a). Supporting data are presented and discussed point by point in the following sections. As will be explained, we concluded that [Fe]/[Se]=1.1 and $T_s$=500°C are the optimum conditions with respect to both the crystallinity and surface flatness of the FeSe thin films. Epitaxial growth of FeSe on STO single crystals was observed over a wide $T_s$ range from 400–600°C and a wide [Fe]/[Se] region from 0.7–1.25 around the optimum conditions (area with purple filled squares). In the Fe-poor and high-$T_s$ region, the impurity phase $Fe_7Se_8$, which has a quite different crystal structure to that of tetragonal FeSe, segregated (open triangles), which was also reported in Ref. [38]. Note that $Fe_7Se_8$ and other impurity phases were not detected in the films grown at the optimum $T_s$. In the Fe-rich region, only Fe metal was detected as an impurity phase (orange filled triangles). Previous reports on bulk FeSe synthesized at 680°C indicated that the Fe and FeSe phases clearly separated at high [Fe]/[Se] because the solubility limit of Fe in FeSe is as low as ~5% [31]. Conversely, phase separation was not detected in our films up to an excess Fe concentration of ~25%, indicating that the nonequilibrium MBE growth process tolerates much higher Fe solubility without segregation.

Figure 1(b) illustrates the influence of introduced strain on the FeSe thin films. The FeSe film with [Fe]/[Se]=1.1 (i.e., close to stoichiometric FeSe) is under tensile and compressive strain along the in-plane and out-of-plane directions, respectively, because the in-plane lattice parameter of the STO substrate is larger than that of FeSe. Conversely, in FeSe with [Fe]/[Se]>1.1 (i.e., Fe-rich chemical composition), strain is introduced only in the in-plane direction; the out-of-plane lattice parameter is almost relaxed. Resulting from these different strain structures, the FeSe films with [Fe]/[Se]≤1.1 exhibit lower activation energies for electrical conductivity than those of films with [Fe]/[Se]>1.1, even though insulator-like behavior is observed for all the obtained films. This behavior will be discussed in detail in Secs. III-4 and III-5.

*III-2. Optimization of $T_s$*

Here, we examine in detail the relationships between deposition conditions, crystalline phases, orientations, and crystallinity according to analysis of raw experimental data. After preliminary experiments, the ratio of Fe flux rate to Se flux rate was fixed at ~1:10, for which the temperatures of the Fe and Se cells were 1110 and 135°C, respectively. Because of the much higher evaporation rate of Se than Fe during film growth, we used an Se flux rate that was ten times higher than that of Fe. Figure 2(a) shows out-of-plane XRD patterns of the FeSe films grown at $T_s$ of 350–700 °C. The thicknesses of the films grown at $T_s$ of ~350°C (corresponding growth rate of ~2 Å/min) and ≥470°C (~3 Å/min) were ~15 and ~20 nm, respectively. At $T_s$>600°C, Fe impurities and misoriented FeSe crystallite diffraction peaks such as FeSe 101 were detected, which is consistent with Ref. [18]. An unidentified impurity phase [indicated by the solid circle in Fig. 2(a)] segregated at the lowest $T_s$ examined (350°C). Although we speculate that this impurity phase is most plausibly $Fe_3Se_4$, we could not confirm this because other diffractions, such as FeSe 110, possibly appear at almost the same $2\theta$ angles. Single-phase *c*-axis-oriented FeSe films were obtained only at $T_s$=470 and 530 °C (i.e., at ~500 °C). At these $T_s$, the obtained [Fe]/[Se] was ~1.0 (i.e., stoichiometric) because of the extremely high Se flux rate. Figures 2(b)–(d) show the surface morphology of the films grown at $T_s$=700, 530, and 350°C, respectively. The large grains with lateral sizes of 80–300 nm and heights of 30–60 nm from the FeSe surface in Fig. 2(b) were attributed to impurity Fe, which was detected at $2\theta$=44.6° (Fe 110 diffraction) in the XRD pattern in Fig. 2(a) and confirmed with an FE-SEM/EDX point analysis. Small particles with lateral sizes of ~40 nm and heights of ~10 nm in Fig. 2(c) originated from surface degradation of the samples induced by air exposure during AFM observation [37]. For all the films, we evaluated the crystallinity along the *c* axis from $\Delta\omega$ of the FeSe 001 diffraction and surface roughness from $R_{rms}$. Figure 2(e) plots these values as a function of $T_s$. The minimum $\Delta\omega$ (0.05° at $T_s$=530°C) and $R_{rms}$ (1.0 nm at $T_s$=470°C) are both located at $T_s$ of ~500 °C, indicating that 500°C is the optimum $T_s$ for FeSe film growth. Except at the optimum $T_s$, the segregation of the misoriented FeSe and impurity crystallites prevents lateral growth of *c*-axis-oriented FeSe domains and, consequently, $\Delta\omega$ gradually increases ($\Delta\omega$ = 0.26° for $T_s$=350°C and 0.19° for $T_s$=700°C). Additionally, the formation of large Fe grains at higher $T_s$ and the suppression of lateral migration at lower $T_s$ both lead to inhomogeneous surfaces, as observed in Figs. 2(b) and 2(d), and increasing $R_{rms}$ with elevating $T_s$ ($R_{rms}$=5.3 nm for $T_s$=350°C and $R_{rms}$=20 nm for $T_s$=700°C). Consequently, both $\Delta\omega$ and $R_{rms}$ exhibit the inverted bell-shaped curves with respect to $T_s$, as seen in Fig. 2(e).

*III-3. Optimization of [Fe]/[Se]*

Next, we optimized the chemical composition of the films by the varying the flux rates of Fe and Se, which were controlled independently of the temperature of each Knudsen cell. The Fe and Se cells were heated at 1050–1100 and 127.5–140 °C, respectively. In these cell-temperature ranges, flux pressures were varied between $3\times10^{-6}$–$3\times10^{-5}$ Pa for Fe and $2\times10^{-5}$–$2\times10^{-4}$ Pa for Se. Here, the target film thickness was ~10 nm, which was exactly determined by XRR and interference detected in out-of-plane XRD measurements, to avoid lattice relaxation and effectively introduce film strain. The other growth parameters were all fixed (e.g., the optimum $T_s$=500°C obtained in the previous section). Figure 3(a) shows out-of-plane XRD patterns of the films with [Fe]/[Se] of 0.7–1.9. All the films exhibited clear *c*-axis orientation in the out-of-plane direction. In the films with [Fe]/[Se]=1.3 and 1.9, small amounts of an Fe impurity were observed because the Fe flux rate was too high. Figures 3(b) and 3(c) show RHEED patterns of the films with [Fe]/[Se]=1.3 and 1.0, respectively. Clear streak patterns were observed in the RHEED patterns for both films, confirming their heteroepitaxial growth. As indicated by the red arrows in Fig. 3(c), unidentified streak diffractions appeared, which are attributed to surface reconstruction within the shallow penetration depth of the incident electron beam (<1 nm) [39]. In-plane growth along the STO in-plane lattice (i.e., [100] FeSe∥[100] STO) and fourfold in-plane symmetry for the 200 diffraction originating from the tetragonal lattice of FeSe were detected for all the films (see Fig. S1 in the Supplemental Material [40] for in-plane diffraction patterns), revealing that all the FeSe films grew heteroepitaxially on STO (001) substrates without any rotational domains at the optimum $T_s$ of 500°C irrespective of chemical composition.

The crystallinities along the $c$ and $a$ axes are plotted as a function of chemical composition in Fig. 3(d) (see Fig. S2 in the Supplemental Material [40] for raw diffraction patterns). These results suggest that chemical compositions close to stoichiometry can grow highly crystalline FeSe epitaxial films with small orientation twist and tilt angles along the in-plane direction ($\Delta\phi$=0.4° for films with [Fe]/[Se]=0.9 and 1.1) and out-of-plane direction ($\Delta\omega$=0.009° for the film with [Fe]/[Se]=1.1). We note that all the out-of-plane FWHMs (in the broadest case, $\Delta\omega$=0.054° for the film with [Fe]/[Se]=0.8) are narrower than those of previously reported films ($\Delta\omega$=0.2° for a ~200-nm-thick FeSe film grown on an LaAlO$_3$ substrate via pulsed laser deposition [41]), which is probably because of the atomically flat surface of the pre-treated STO substrates and slow growth rate in our MBE process. Figures 3(e)–3(i) show the surface morphologies of FeSe films with [Fe]/[Se]=0.7–1.9 (that with [Fe]/[Se]=1.1 was reported in Ref. [21]). The films with [Fe]/[Se]=1.0−1.3 have almost no surface pits, whereas many pits were observed in the films with [Fe]/[Se]=0.7, 0.8, and 1.9. In the films with low [Fe]/[Se] (0.7 and 0.8), we consider that the origin of the pits is evaporation of excess Se. This would be similar to the case of GaAs [42, 43], where oxide desorption from the GaAs surface generates surface pits. Similarly, excess Se prevents lateral growth of FeSe domains because the excess Se forms crystals on the substrate surface, which subsequently reevaporate as the gas phase at $T_s$=500°C because $T_s$ is higher than the Se evaporation temperature (<150°C). In the Fe-rich region, FeSe and Fe crystals coexist at the beginning of nucleation, which also suppresses lateral migration of FeSe deposition precursors. Corresponding to the generation of surface pits, the $R_{rms}$ values of films with [Fe]/[Se]=0.7, 0.8, and 1.9 in Fig. 3(j) were larger than those of films with [Fe]/[Se]=0.9–1.3. The flattest surface was obtained for the FeSe film with [Fe]/[Se]=1.1 ($R_{rms}$=0.6 nm) [21] and the roughest surface with $R_{rms}$=5.0 nm was observed for the film with [Fe]/[Se]=0.7. From these results, we concluded that [Fe]/[Se]=1.1 is the optimum composition with respect to both crystallinity and surface flatness.

*III-4. Effects of lattice strain on FeSe film properties*

Figure 4(a) illustrates the variations of the *a*- and *c*-axis lattice parameters of the FeSe films with [Fe]/[Se]=0.7–1.9 deposited at $T_s$=500 °C, which were estimated from the FeSe in-plane 200 diffraction and using the Nelson-Riley function [44] for out-of-plane 00*l* peaks, respectively [see Fig. S1(a) in the Supplemental Material [40] for in-plane diffraction patterns and Fig. 3(a) for out-of-plane XRD data]. There is a large in-plane lattice mismatch between tetragonal FeSe and the STO substrate; i.e., the in-plane lattice parameter of the STO substrate ($a_{sub}$=3.905 Å) is larger than that of bulk Fe$_{1.09}$Se ($a_{bulk}$=3.7734 Å when [Fe]/[Se] is ~1.1 [30]), which gives a mismatch of ($a_{sub}$−$a_{bulk}$)/$a_{sub}$=3.4%. As a result, the *a*- and *c*-axis lattice parameters of the FeSe film with [Fe]/[Se]=1.1 were expanded to 3.836 Å by tensile strain along the *a* axis, and compressed to 5.455 Å by compressive strain along the *c* axis, respectively. For the films at both ends of the chemical compositions, the lattice parameters were *a*=3.795 Å and *c*=5.530 Å for the film with [Fe]/[Se]=0.7 and *a*=3.824 Å and *c*=5.533 Å for the film with [Fe]/[Se]=1.9. The maximum *a* and minimum *c* values were observed for the film with [Fe]/[Se]=1.1, which is close to stoichiometric value ([Fe]/[Se] = 1.0). Note that the lattice parameters of bulk FeSe vary with [Fe]/[Se]. In the region where [Fe]/[Se]<1.0 (Se rich), it is known that Fe-vacancy sites appear in the FeSe structure. Although the vacancy sites are usually randomly distributed (i.e., disordered), it has been reported that ordered states are stabilized via hydrothermal or high-pressure synthesis processes [33]. The Fe-vacancy ordering (i.e., emergence of superlattice diffractions in XRD) shrinks the *a*-axis lattice parameter ($a$=3.67(1) Å and $c$=5.70(3) Å for $Fe_{0.75}Se$ [33]) and expands the *c*-axis one. For films with chemical compositions ranging from [Fe]/[Se]=0.7–1.0, the trend of the lattice parameter variation for *a* and *c* axes is similar to that for bulk FeSe, as shown by the red filled circles and blue filled squares in Fig. 4(a), respectively, although the absolute values between the present epitaxial films and bulk samples for each chemical composition are different because of the introduced strain in the former. Because no superlattice diffraction peak was detected in our films with [Fe]/[Se]≤1.0, we speculate that Fe vacancies were disordered or the volume fraction of the ordered superstructure was too small to detect by XRD.

In the [Fe]/[Se]≥1 region, both the *a*- and *c*-axis lattice parameters of bulk scarcely changed despite the increase of excess Fe concentration ($a$=3.7692 Å and $c$=5.5137 Å in $Fe_{1.33}Se$ [31]) compared with those of stoichiometric FeSe ($a$=3.7720 Å and $c$=5.5161 Å [31]) because of the low solubility limit of Fe (i.e., almost all excess Fe was segregated as impurity Fe). However, in our films, lattice parameters varied dramatically with increasing [Fe]/[Se], suggesting that the excess Fe exceeding the equilibrium solubility limit was incorporated into the FeSe lattice, which was probably facilitated by the nonequilibrium film growth process. Thus, there are two possibilities for the origin of the change in lattice parameters of the films with composition, i.e., mechanical strain introduced from the substrate and chemically induced strain caused by varying the chemical composition. Because we could not exactly distinguish these two factors from the variations of the *c*- to *a*-axis ratio (*c*/*a*) and the volume of the structure (see Fig. S3 in the Supplemental Material [40] for raw data), we defined the changes of the lattice parameters from stoichiometric bulk FeSe as measures of the introduced strain $\Delta a=(a_{film}-a_{bulk})/a_{bulk}$ and $\Delta c=(c_{film}-c_{bulk})/c_{bulk}$, where $a_{bulk}$=3.7735 Å and $c_{bulk}$=5.5238 Å when [Fe]/[Se]=1.0 [30]. These parameters are plotted in Fig. 4(b) as a function of [Fe]/[Se]. The change ratios for the optimum composition of [Fe]/[Se]=1.1 are $\Delta a$=+1.66% and $\Delta c$=−1.24%. In the region with [Fe]/[Se]<1.0 (Se rich), $\Delta a$ decreases and $\Delta c$ increases with decreasing [Fe]/[Se] from stoichiometric ($\Delta a$=+1.37% and $\Delta c$=−0.99%) to the lowest ratio ($\Delta a$=+0.56% and $\Delta c$=+0.11%). In the [Fe]/[Se]≥1 region, $\Delta a$ decreases and $\Delta c$ increases with increasing [Fe]/[Se]. Because the chemically induced strain originates from Fe vacancies in films with [Fe]/[Se]<1.0 and excess Fe in the crystal lattice in films with [Fe]/[Se]>1, $\Delta a$ and $\Delta c$ exhibit the bell-shaped curves.

In the FeSe film with [Fe]/[Se]=1.9, the *c*-axis lattice parameter almost relaxed ($\Delta c$=+0.16%) to that of the bulk and the *a*-axis lattice parameter remained under tensile strain ($\Delta a$=+1.34%) like the optimally grown FeSe film with [Fe]/[Se]=1.1 ($\Delta a$=+1.66%), although slight relaxation occurred. We consider that the origin of this uniaxial relaxation is mainly the excess Fe incorporated in the FeSe lattice, although the site of the excess Fe could not be precisely determined. A possible site of the stabilized excess Fe inside the lattice is the 2*c* site (represented by multiplicity and Wyckoff letter) in the tetragonal *P*4/*nmm* (No. 129) structure (Fig. S4 in the Supplemental Material [40] for crystal structures), similar to the analogous iron chalcogenide tetragonal $Fe_{1+y}Te$ [45, 46]. Excess Fe should expand the *c*-axis lattice parameter because it increases the repulsion between layers and/or neutralizes the charge of the top and bottom Se layers, which should be achieved by the non-equilibrium growth process in the Fe-rich flux atmosphere. These factors would provoke the anisotropic relaxation of the crystal lattice. Because the relaxation trends of the FeSe lattice are different in the regions with low and high [Fe]/[Se], $\Delta a$ vs $\Delta c$ relations are re-plotted in Fig. 4(c), giving two straight lines for [Fe]/[Se]=0.7–1.1 (slope of ~1.2) and for 1.1–1.9 (slope of ~4.2). The different slopes imply that different strain effects are present in the two regions. The electronic structure should also be different between the regions with low and high [Fe]/[Se] because the electronic structure of FeSe is quite sensitive to structural changes [14−16].

*III-5. Electronic transport properties and electronic structures*

To investigate the two different strain effects seen in Fig. 4(c) on the electronic transport properties of the sample, $\rho-T$ relationships were examined. Figure 5(a) shows $\rho-T$ curves of the films with [Fe]/[Se]=0.8–1.9 grown at the optimum $T_s$ of 500°C. For the two Fe-rich films (i.e., [Fe]/[Se]=1.3 and 1.9), insulator-like behavior (i.e., increasing $\rho$ with decreasing $T$) was observed, and the increase of $\rho$ was particularly enhanced at <50 K. The films with [Fe]/[Se] ratios of ≤1.1 also exhibited insulator-like behavior, especially at lower temperatures (<30 K), as shown in Fig. 5(b). To clarify differences in these insulator-like behaviors, we constructed Arrhenius plots with the proportional relation of the logarithm of electrical conductivity ($\sigma=1/\rho$) to $T^{-1}$ in Figs. 5(c) and S5(a) (see Supplemental Material [40] for raw plots) (i.e., $\sigma = \sigma_0 exp(-E_a/k_B T)$, where $E_a$, $\sigma_0$, and $k_B$ are the activation energy, a prefactor constant, and the Boltzmann constant, respectively) and to $T^{-1/4}$ for the films with [Fe]/[Se]≤1.1 in Fig. 5(d), ≥1.3 in Fig. 5(e), and the whole [Fe]/[Se] range in Fig. S5(b) (see Supplemental Material [40] for raw plots) [corresponding to the Mott's variable-range hopping model [47], $\sigma = \sigma_0 exp(-(T_0/T)^{1/4}$, where $\sigma_0$ and $T_0$ are a constant and a hopping parameter, respectively). $T^{-1/4}$ was used because the fitting result obtained by $T^{-1/4}$ (i.e., three-dimensional) was the best among the fitting results with $T^{-1/n}$, where $n$=2–4 (see Fig. S6 of the Supplemental Material [40] for raw plots and Table S1 [40] for fitted results); i.e., the minimum standard deviation was obtained in the case of $T^{-1/4}$ fitting. Linear proportional relationships between $\sigma$ and $T^{-1}$ were clearly observed for all the films in the high-$T$ region, but not in the low-$T$ region. In contrast, the $\sigma-T^{-1/4}$ plots also presented straight lines in the low-$T$ region. From these relationships, $E_a$ and $T_0$ were estimated for each film and are plotted in Fig. 5(f). Both $E_a$ and $T_0$ increased discontinuously for films with [Fe]/[Se] above and below 1.1, indicating that the electronic structures in films with [Fe]/[Se]≥1.3 ($E_a$=23 meV and $T_0$=~1×10$^6$ K for the film with [Fe]/[Se]=1.9) were quite different from those in [Fe]/[Se]≤1.1 ($E_a$=1.2 meV, $T_0$=~100 K for the film with [Fe]/[Se]=0.8, and $E_a$=1.1 meV and $T_0$=~13 K for the film with [Fe]/[Se]=1.1). The transition composition corresponds well to that of the strain effect boundary in Fig. 4(c) (i.e., [Fe]/[Se]=1.1), which is the best composition with respect to both crystallinity and surface flatness [see Figs. 3(d) and 3(j)]. It should be noted that the films with the boundary composition (i.e., [Fe]/[Se]=1.1) exhibited the highest $T_c$ of 35 K when we applied a gate bias to the EDLT in Ref. [21], whereas the samples grown at growth rates that were too high ($E_a$=0.1 meV and $T_0$=0.002 K) and low ($E_a$=1.2 meV and $T_0$=16 K) exhibited quite broad superconducting transitions without zero resistance and lower $T_c$ [22].

The electronic transport properties in Fig. 5 revealed that all the fabricated FeSe thin films exhibited up-turn $\rho$–$T$ relationships, which are usually interpreted as typical of electrical insulators. Here it should be noted that such up-turn behavior of $\rho$−$T$ is observed even for metals because of magnetic impurity scattering known as the Kondo effect. However, the relationship of linear $\rho$ vs. ln $T$ at low $T$, which is consistent with the Kondo effect, was not observed for our FeSe films (see Fig. S7 of the Supplemental Material [40] for raw plots), indicating that the observed insulator-like behavior should not be related to the Kondo effect. The possible origins of the up-turn behavior include (i) the change in the electronic structure from metallic to insulating (i.e., open band gaps) and/or (ii) scattering of conducting carriers (e.g., potential barrier). Therefore, we performed ARPES measurements to directly examine whether the FeSe films had open band gaps. Fabricated thin FeSe films were transferred *in situ* from the growth chamber to the ARPES measurement chamber under an ultrahigh vacuum of <1×10$^{-7}$ Pa to measure the electronic structure without any surface pretreatment such as thermal annealing or sputtering. Figures 6(a) and 6(b) show ARPES spectra measured at ~10 K for the optimally grown FeSe film with [Fe]/[Se]=1.1 around the Γ and M points, respectively. To clearly visualize the band dispersion around $E_F$, second-derivative spectra with respect to energy around the Γ and M points are presented in Figs. 6(c) and 6(d), respectively. At the Γ point, a hole-like band derived from Fe 3$d$ orbitals [48] intersected $E_F$ and its top was located at a binding energy of $E_b$−$E_F$=−15 meV. At the M point, an electron-like band crossed $E_F$ and its bottom was located at $E_b$−$E_F$=+5 meV. These results confirm that the electronic structure in the insulator-like FeSe is metallic (i.e., the band gap is not opened at $E_F$) even though separated hole- and electron-like bands exist around $E_F$ (i.e., multiband structure), like bulk FeSe [49]. Here, it should be noted that their energy overlap (i.e., the difference of $E_b$ between the top of the hole-like band at Γ and the bottom of the electron-like band at M) is as small as ~20 meV, which is the same value as that determined by ARPES for FeSe with in-plane tensile lattice strain [50]. Because a small energy overlap results in enhancement of nematic instability [50], the lack of superconductivity in all our FeSe films would also originate from nematic instability.

Because superficially antagonistic results were obtained for the electronic transport properties and ARPES measurements (i.e., a linear proportional relation between ln $\sigma$ and $T^{-1/4}$ despite the metallic electronic structure), we performed Hall-effect measurements for two representative films in the two regions; that is, [Fe]/[Se]=1.1 (the most strained and almost stoichiometric film) and 1.9 (the most Fe-rich one). Figures 7(a) and 7(b) show the magnetic field ($H$) dependences of transverse resistivity ($\rho_{xy}$) for the films with [Fe]/[Se]=1.1 and 1.9, respectively. Because the slopes of the plots in Figs. 7(a) and 7(b) are both positive, the dominant carrier type in the two films at room temperature is holes, which is similar behavior to that reported in Ref. [51]. The dominant carrier changed to electrons around 120 and 170 K for the films with [Fe]/[Se]=1.1 and 1.9, respectively, where linear relations of $\rho_{xy}$ against $H$ were not observed, presumably because the number of hole carriers became comparable to that of electrons and Hall voltages were compensated. From 300 K to the lowest temperature where definite Hall voltages were observed, we estimated the Hall coefficient ($R_H$), which is presented in Fig. 7(c). Although the $R_H$–$T$ behavior of both films was roughly similar, i.e., $R_H$ increased and then decreased with decreasing temperature, $R_H$ for the film with [Fe]/[Se]=1.9 was ca. two orders of magnitude higher than that for the film with [Fe]/[Se]=1.1 over the whole $T$ range. Carrier concentration ($n$) was estimated based on the single carrier model relation of $1/q|R_H|$, where $q$ is elementary charge, irrespective of the multiband electronic structure observed in the ARPES measurements (Fig. 6). We found that $n$ in the film with [Fe]/[Se]=1.1 was ~2×10$^{21}$ cm$^{-3}$ at 300 K, which is approximately one order of magnitude higher than that of bulk FeSe (~3×10$^{20}$ cm$^{-3}$) [52] but lower than that of superconducting thin films with $T_c$=11.4 K, where $R_H$=~0.5 cm$^{-3}$/C at 300 K ($n$=~10$^{22}$ cm$^{-3}$) [53]. Meanwhile, $n$ in the film with [Fe]/[Se]=1.9 was two orders of magnitude lower (~1×10$^{20}$ cm$^{-3}$ at 300 K) than that in the [Fe]/[Se]=1.1 sample. This difference should arise from ionic Fe$^{2+}$ inside the lattice, which acts as the electron donor via isovalent doping [54], in the film with [Fe]/[Se]=1.9.

Next, we estimated the mobility ($\mu$) of the two films from the inverse of $\rho$ in Fig. 5(a) and estimated $n$. $\mu$ at 300 K were ~3×10$^0$ and ~4×10$^{-3}$ cm$^2$/(Vs) in the films with [Fe]/[Se]=1.1 and 1.9, respectively; these values are respectively one and four orders of magnitude smaller than that of bulk FeSe (~25 cm$^2$/(Vs) at 300 K [51]). Considering the relationship between $\sigma$ and $n$, the difference in $\mu$ should be strongly related to the different $E_a$ in the insulator-like behavior of the two thin films. $\mu$ in the films with [Fe]/[Se]=1.1 and 1.9 decreased to 8×10$^{-1}$ and 1×10$^{-3}$ cm$^2$/(Vs), respectively, at the lowest temperatures (150 and 200 K, respectively). This indicates that there is a potential barrier derived from conduction carrier scattering. To unveil the potential barrier, we plotted ln($\mu$) versus $T^{-1}$ in Fig. 7(d). This plot corresponds to the percolation conduction model with distributed potential barriers, $\mu(T)=\mu_0\exp(-e(\Phi_0 - (e\sigma^2_\Phi/2k_BT))/k_BT)$ [55, 56], where $\mu_0$, $\Phi_0$, and $\sigma_\Phi$ denote a virtually $T$-independent constant, average potential barrier, and the distribution width of the potential barrier, respectively. The solid curves in Fig. 7(d) represent the fitting results obtained using the above equation, which reproduce the experimental results well. The calculated $\Phi_0$ and $\sigma_\Phi$ values were ~77 and ~43 meV for the film with [Fe]/[Se]=1.1 and ~150 and ~74 meV for that with [Fe]/[Se]=1.9, respectively. Even the film with [Fe]/[Se]=1.1 has high and wide potential barriers, like disordered oxide semiconductors [55, 56]. Thus, we concluded that the insulator-like behavior of the very thin FeSe epitaxial films is because carriers cannot move freely due to the influence of potential barriers in the conduction band despite the closed-band-gap metallic electronic structure. Such a high potential barrier is tentatively attributed to the presence of large amounts of excess Fe. Moreover, the potential barrier in the FeSe film with [Fe]/[Se]=1.9 was much higher than that in the film with [Fe]/[Se]=1.1, which caused $E_a$ and $T_0$ to increase.

Finally, we discuss the origin of the potential barrier resulting from structural variation caused by excess Fe and introduced strain. Even though the FeSe film with higher $\mu$ ([Fe]/[Se]=1.1) shrinks along the $c$ axis, that with lower $\mu$ ([Fe]/[Se]=1.9) relaxes to resemble a bulk state [see Fig. 4(a)], indicating that two dimensionality is enhanced along the in-plane direction in the film with [Fe]/[Se]=1.9. Thus, the structural change in [Fe]/[Se]=1.9 not only suppresses conduction along the in-plane direction, but also decreases the dimensionality in the out-of-plane direction, causing $\mu$ to lower dramatically (i.e., the potential barrier becomes higher) compared with that in the film with [Fe]/[Se]=1.1. Overall, we succeeded in tuning the electronic transport properties of insulator-like FeSe through structural variation stimulated by introducing excess Fe and lattice strain via a nonequilibrium MBE film growth process.

## IV. Conclusion

We fabricated ~10 nm-thick insulator-like FeSe films with [Fe]/[Se]=0.8–1.9 at the optimized growth temperature of 500 °C by MBE and investigated their electronic transport properties and electronic structure. It was found that the lattice strain introduced in the films exhibited unusual behavior; the almost stoichiometric film with [Fe]/[Se]=1.1 had the largest in-plane tensile strain, whereas the lattice parameters approached the unstrained values for the films with [Fe]/[Se] values that were nonstoichiometric. In the films with [Fe]/[Se]<1.1 (Se-rich region), the FeSe lattice contracted in plane and expanded perpendicular to the substrate, probably because of the introduction of disordered Fe vacancies. In the Fe-rich region, both the $a$ and $c$ axes were expanded from the unstrained values, which originated from the excess interstitial Fe stabilized by the nonequilibrium MBE film growth process. $\rho$–$T$ measurements and electronic structure observations provide apparently contradicting results. We found that all films exhibited insulator-like behavior, whereas ARPES measurements indicated that the films had metallic electronic structures. The insulator-like behavior was classified into two regions according to $E_a$ with the boundary composition of [Fe]/[Se]=1.1; $E_a$ of the Fe-rich films were one order of magnitude higher than those of the Se-rich films. $\mu$ of the [Fe]/[Se]=1.1 film with small $E_a$ was three orders of magnitude higher than that of the Fe-rich film with large $E_a$, which originated from the potential barrier distributed in the conduction band. Such a high potential barrier is tentatively attributed to the presence of large amounts of excess Fe. These findings explain the above contradicting results of the coexistence of metallic electronic structure and insulating electrical properties. The present results may provide an explanation for the previously reported broad superconducting transition without zero resistance under an applied gate bias in an EDLT with film formed at high growth rate [22], because high potential barriers would also be induced by the high Fe flux rate.

**ACKNOWLEDGMENTS**

This work was supported by the Ministry of Education, Culture, Sports, Science, and Technology (MEXT) through the Element Strategy Initiative to Form Core Research Center. H. Hiramatsu was also supported by the Japan Society for the Promotion of Science (JSPS) through Grants-in-Aid for Scientific Research (A) and (B) (Grants No. 17H01318 and No. 18H01700), and Support for Tokyotech Advanced Research (STAR).

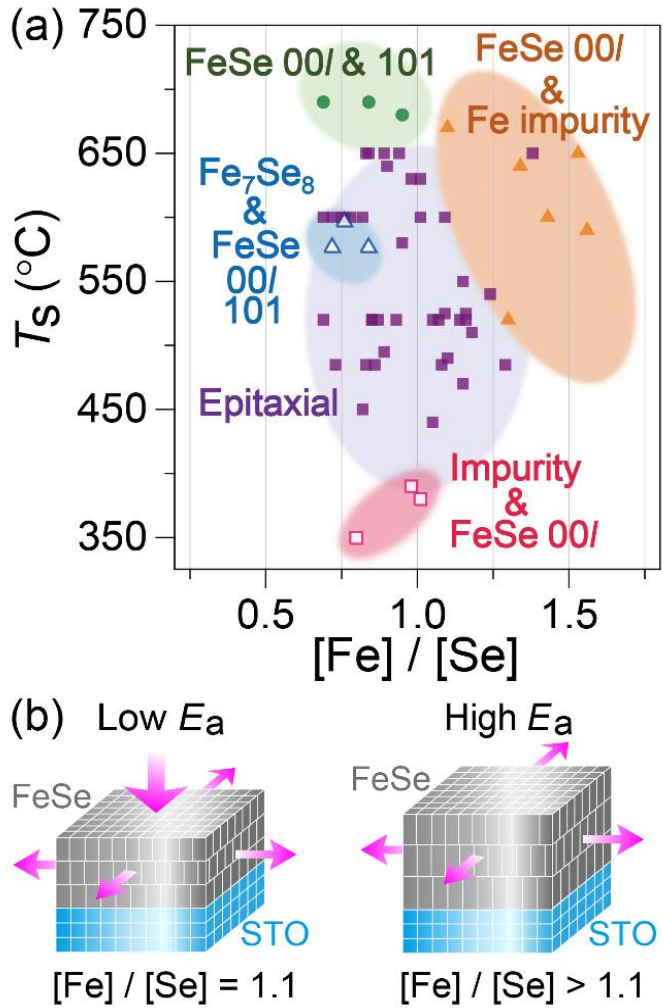

FIG. 1. Overview of FeSe thin films grown by MBE. (a) Relationship between $T_s$ and chemical composition ([Fe]/[Se]) of the fabricated films. Different symbols represent the obtained crystalline phases and orientations. Filled squares denote the epitaxial growth region, open squares the 00*l*-oriented FeSe and an unidentified impurity, closed triangles the 00*l*-oriented FeSe and impurity Fe, open triangles the 00*l*-preferentially oriented FeSe with other orientations and $Fe_7Se_8$, and closed circles the 00*l*-preferentially oriented FeSe with other orientations. (b) Schematics of the strain in FeSe thin films with [Fe]/[Se]=1.1 (left) and >1.1 (right) introduced via thin-film growth. The arrows indicate the directions of the introduced strain. $E_a$ is the activation energy obtained from the electronic transport properties for each strained film.

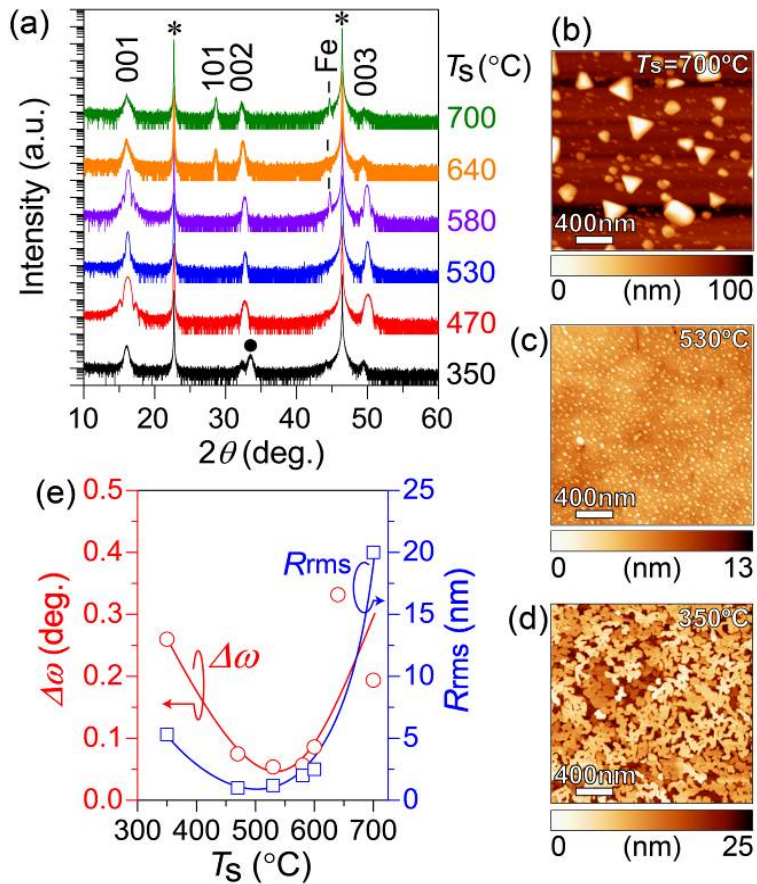

FIG. 2. Structure and surface morphology of FeSe films grown at a constant ratio of Fe flux rate to Se flux rate of ~1:10 and $T_s$=350–700°C. (a) Out-of-plane XRD patterns. The asterisks indicate the substrate diffraction peaks, the solid circle at $2\theta$=33.5° an unidentified phase, and "Fe" at $2\theta$=44.6° iron metal. AFM images of films grown at $T_s$ of (b) 700, (c) 530, and (d) 350°C. Horizontal bars represent height scales. (e) Out-of-plane rocking-curve FWHM for the FeSe 001 diffraction ($\Delta\omega$, red circles) and root-mean-square roughness ($R_{rms}$, blue squares) of the film surface as a function of $T_s$.

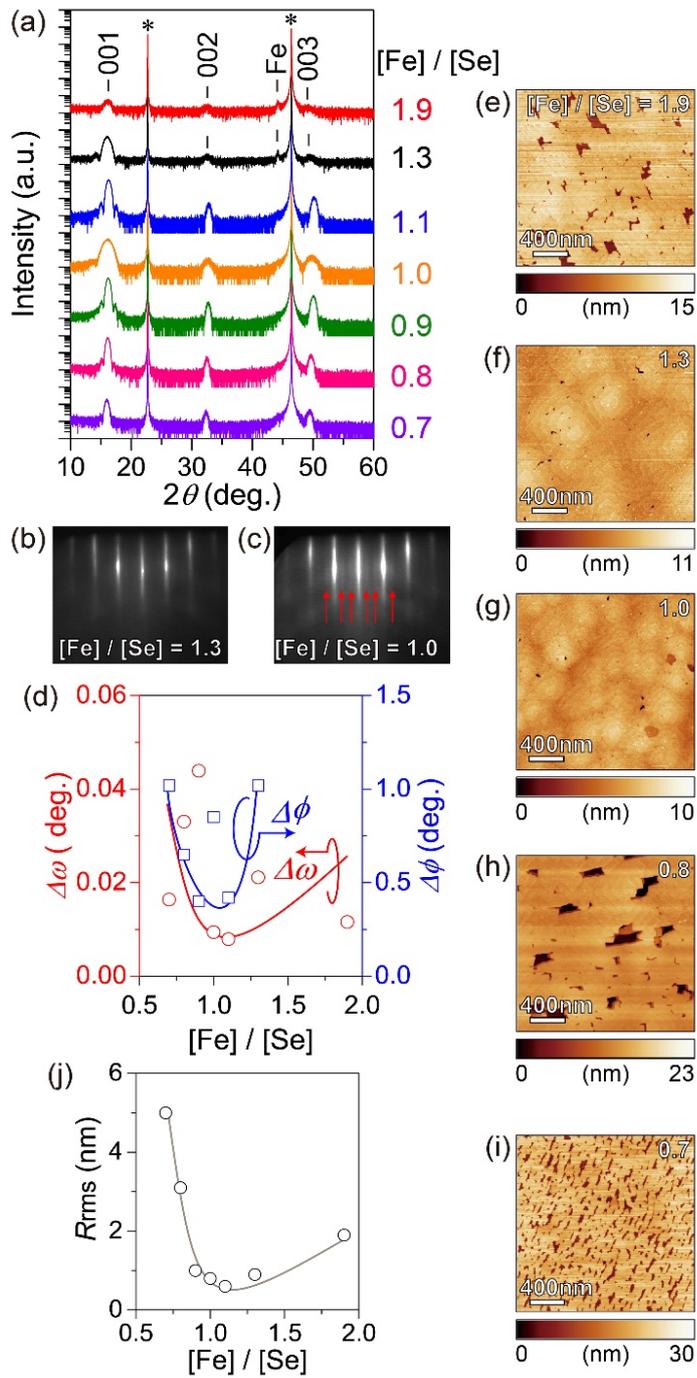

FIG. 3. Quality of FeSe thin films with different chemical compositions ([Fe]/[Se]=0.7–1.9) grown at the optimum $T_s$ of 500 °C. (a) Out-of-plane XRD patterns of the films with different [Fe]/[Se]. The black vertical bars and asterisks denote FeSe 00$l$ diffractions and STO (001) substrate peaks,

respectively. "Fe" indicates the 110 diffraction peak ($2\theta$=44.6°) of impurity Fe, which segregates at [Fe]/[Se]≥1.3. RHEED patterns of FeSe films with [Fe]/[Se]=(b) 1.3 and (c) 1.0. The red vertical arrows in (c) are unidentified diffractions. (d) Rocking-curve FWHMs along the out-of-plane ($\Delta\omega$, red circles) and in-plane ($\Delta\phi$, blue squares) directions taken from the FeSe 001 and 200 diffractions, respectively. (e)–(i) AFM images of FeSe films with [Fe]/[Se]=0.7–1.9 (indicated in each AFM image). An AFM image of FeSe with [Fe]/[Se]=1.1 was reported in Ref. [21]. Horizontal bars represent height scales. (j) Surface roughness ($R_{\text{rms}}$) estimated from (e)–(i) as a function of [Fe]/[Se].

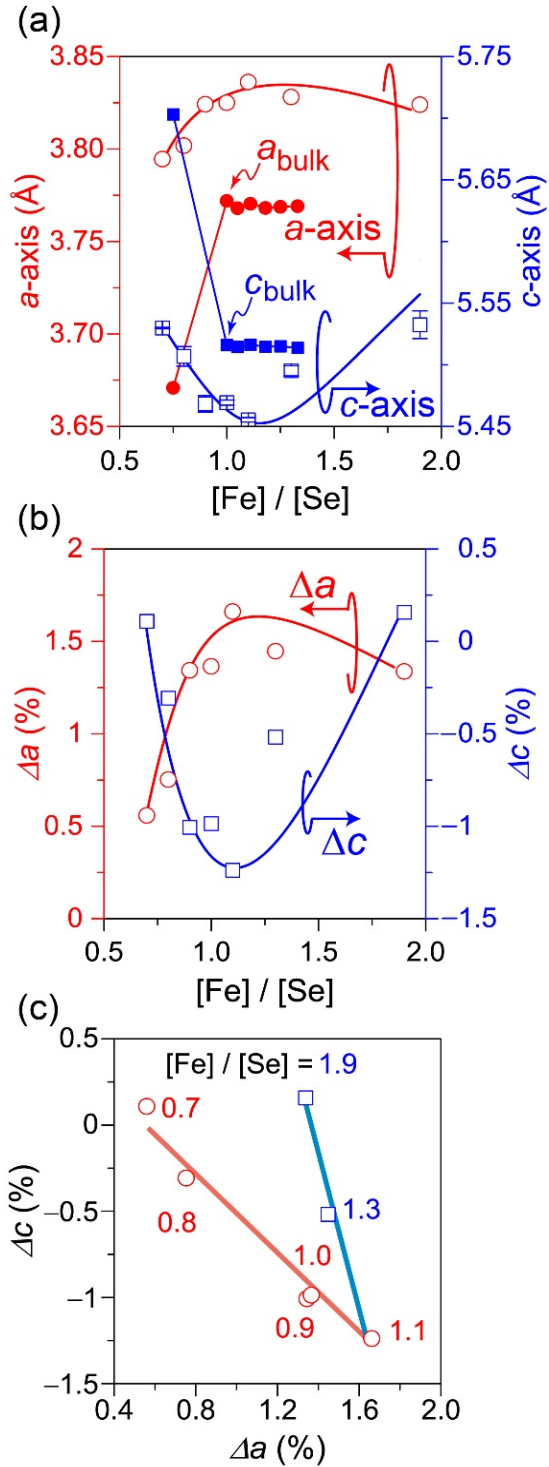

FIG. 4. Relationship between the lattice strain and chemical composition of FeSe films grown at the optimum $T_s$ of 500 °C with [Fe]/[Se]=0.7–1.9. (a) $a$-axis (open red circles) and $c$-axis (open blue squares) lattice parameters. The red closed circles and blue closed squares denote the $a$-axis ($a_{bulk}$) and $c$-axis ($c_{bulk}$) lattice parameters of bulk FeSe [30, 31, 33]. (b) Introduced strain for the $a$ axis (red circles) and $c$ axis (blue squares) calculated by $\Delta a=(a_{film}-a_{bulk})/a_{bulk}$ and $\Delta c=(c_{film}-c_{bulk})/c_{bulk}$ ($a_{bulk}$=3.7735 and $c_{bulk}$=5.5238 Å at [Fe]/[Se]=1.0 [30]). (c) Lattice strain along the out-of-plane direction ($\Delta c$) vs that along the in-plane direction ($\Delta a$). The values in the figure indicate the [Fe]/[Se] values.

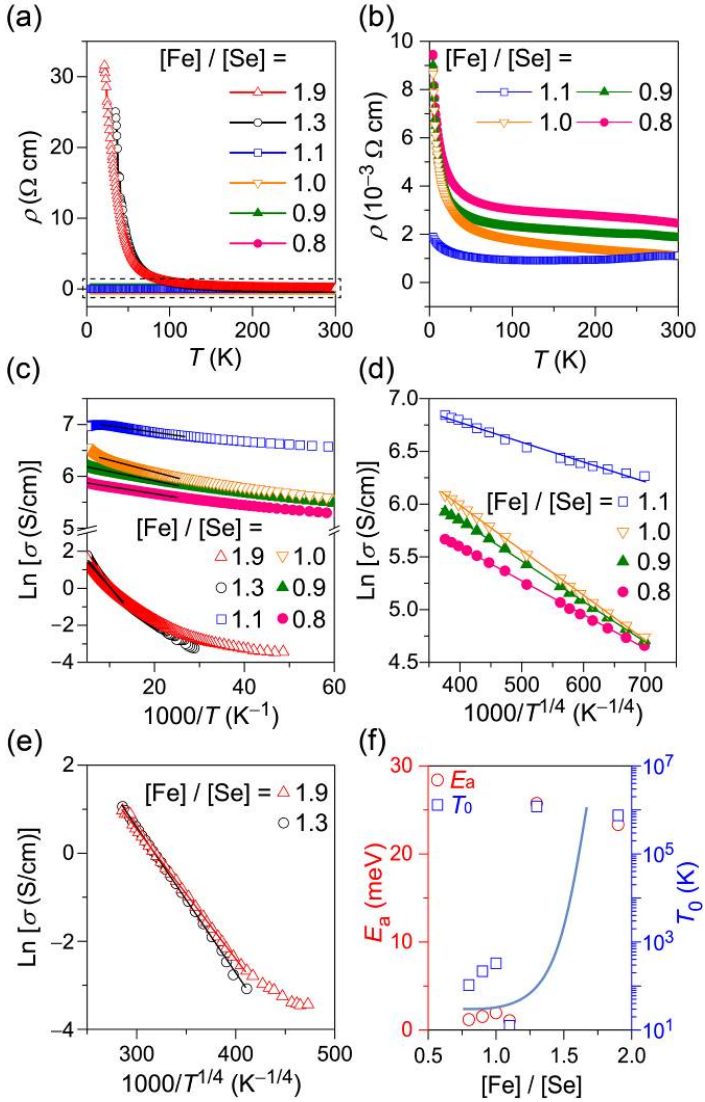

FIG. 5. Electronic transport properties of FeSe films with [Fe]/[Se]=0.8–1.9 grown at the optimum $T_s$ of 500°C. (a) Temperature ($T$) dependence of resistivity ($\rho$). (b) Enlarged $\rho-T$ curves inside the dashed square in (a). (c) Arrhenius plots of electrical conductivity ($\sigma$) against $T^{-1}$. (d), (e) Ln$\sigma$ vs $T^{-1/4}$ plots for the films with [Fe]/[Se] of (d) 0.8–1.1 and (e) 1.3 and 1.9. The straight lines in (c) and (d) are the results of the linear least-squares fitting. (f) Activation energy ($E_a$, circles) estimated from the straight lines in (c) and hopping parameter of the variable-range hopping model ($T_0$, squares) estimated from (d) and (e) as a function of [Fe]/[Se].

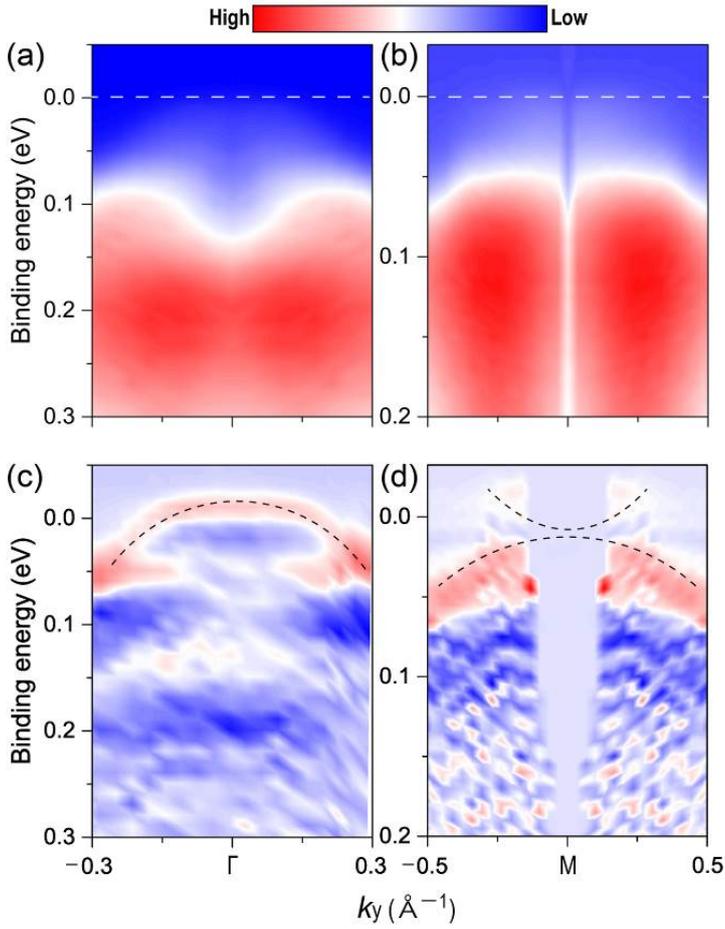

FIG. 6. Fermi surface of the FeSe film grown at the optimum $T_s$ of 500 °C with [Fe]/[Se]=1.1 directly observed by ARPES at ~10 K. The horizontal bar indicates photoelectron intensity. ARPES responses around the (a) Γ and (b) M points. White dashed lines in (a) and (b) represent the Fermi level calibrated using Au as a reference. Second-derivative spectra with respect to energy, where (c) and (d) correspond to (a) and (b), respectively. Black dashed curves in (c) and (d) are a visual guide to help trace the dispersions.

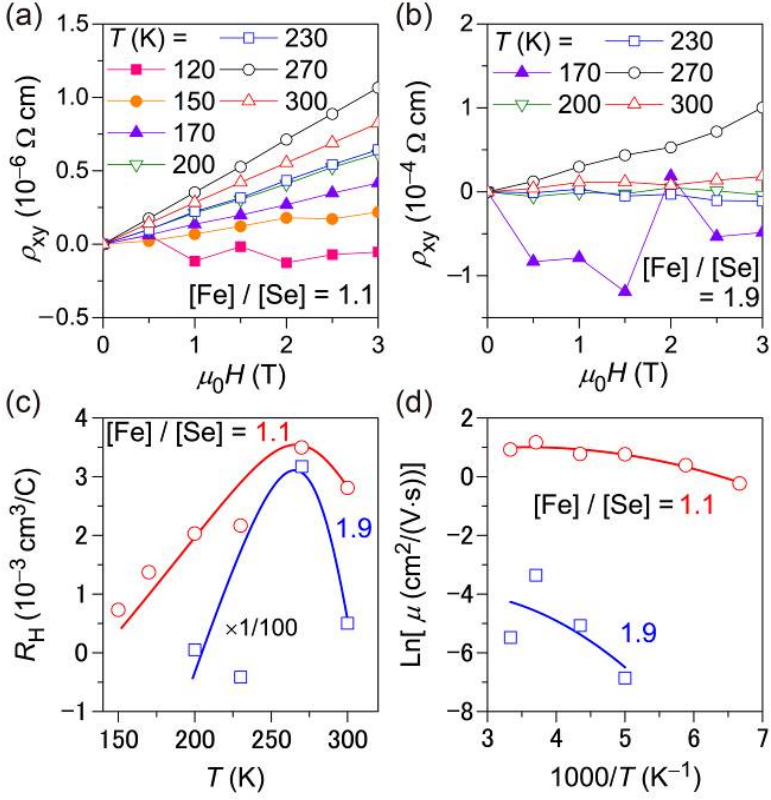

FIG. 7. Results of Hall effect measurements for FeSe films grown at the optimum $T_s$ of 500 °C with [Fe]/[Se]=1.1 and 1.9. (a), (b) Transverse resistivity ($\rho_{xy}$) as a function of external magnetic field ($H$) at (a) $T$=120–300 K for the film with [Fe]/[Se]=1.1 and (b) $T$=170–300 K for the film with [Fe]/[Se]=1.9. (c) Dependence of the Hall coefficient ($R_H$) of the films with [Fe]/[Se]=1.1 and 1.9 on $T$ estimated from (a) and (b), respectively. For the film with [Fe]/[Se]=1.9, the $R_H$ values are magnified by one-hundred times. (d) Dependence of the mobilities ($\mu$) on $T$ estimated from the results in (c) and $\rho$ in Fig. 5(a). Red circles and blue squares in (c) and (d) correspond to the films with [Fe]/[Se]=1.1 and 1.9, respectively. Red and blue solid curves in (d) denote quadratic-polynomial fitting results for the films with [Fe]/[Se]=1.1 and 1.9.

Supplemental Materials for 'Insulator-like behavior coexisting with metallic electronic structure in strained FeSe thin films grown by molecular beam epitaxy'

Kota Hanzawa,[1] Yuta Yamaguchi,[1] Yukiko Obata,[1] Satoru Matsuishi,[2] Hidenori Hiramatsu,[1,2] Toshio Kamiya,[1,2] and Hideo Hosono[1,2]

[1] *Laboratory for Materials and Structures, Institute of Innovative Research, Tokyo Institute of Technology, Mailbox R3-3, 4259 Nagatsuta-cho, Midori-ku, Yokohama 226-8503, Japan*

[2] *Materials Research Center for Element Strategy, Tokyo Institute of Technology, Mailbox SE-1, 4259 Nagatsuta-cho, Midori-ku, Yokohama 226-8503, Japan*

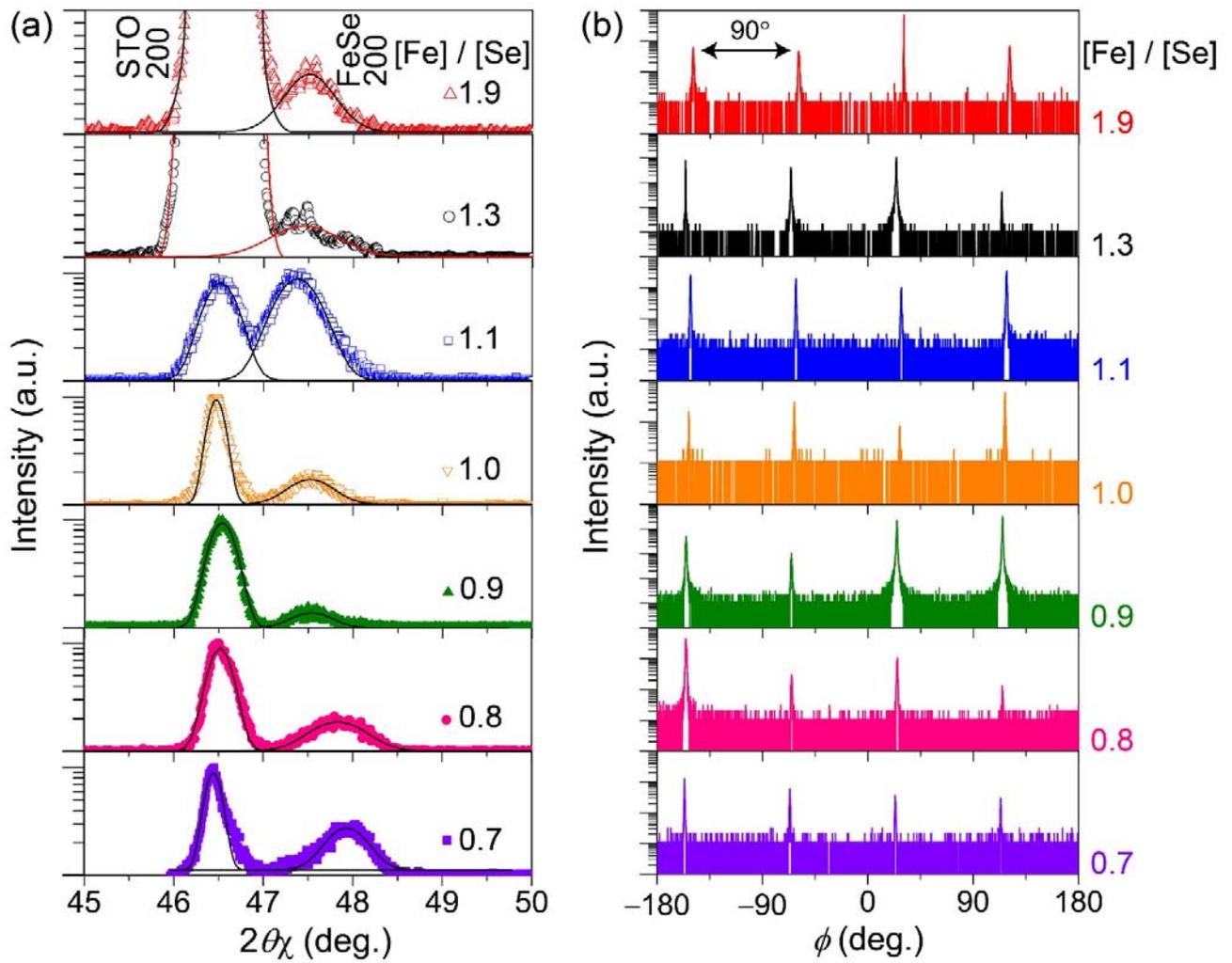

Figure S1. In-plane structure analyses of FeSe thin films grown at 500 °C with [Fe]/[Se]=0.7–1.9. (a) In-plane XRD patterns around the STO 200 diffraction. Solid lines are the fitting results for STO and FeSe 200 diffractions. (b) $\phi$-scans of FeSe 200 diffractions.

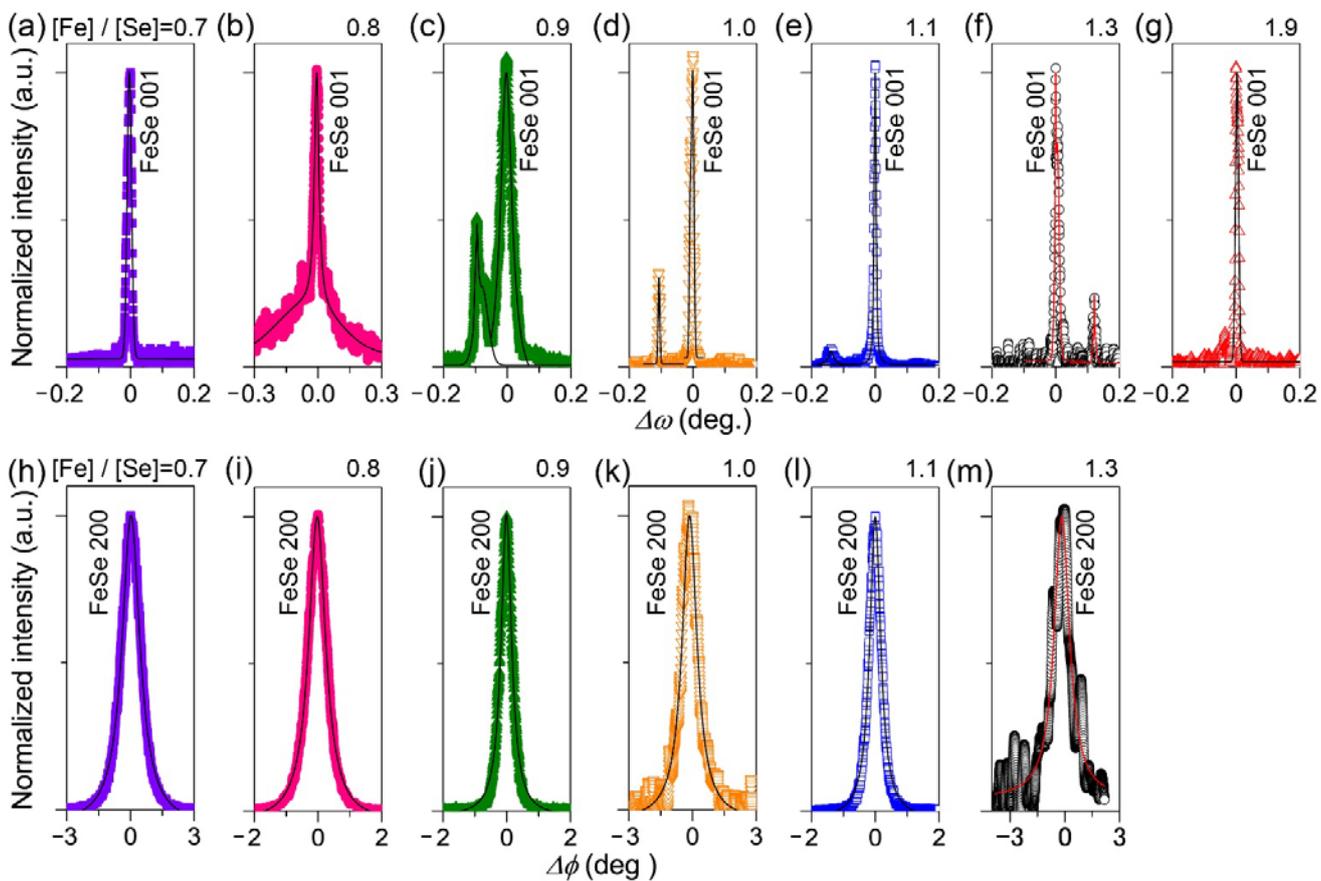

Figure S2. Results of X-ray rocking curve measurements for out-of-plane and in-plane directions. (a)–(g) Rocking curve patterns for the out-of-plane FeSe 001 diffraction of films grown at 500 °C with [Fe]/[Se]=0.7–1.9. (h)–(m) Rocking curve patterns for the in-plane FeSe 200 diffraction of the films with [Fe]/[Se]=0.7–1.3.

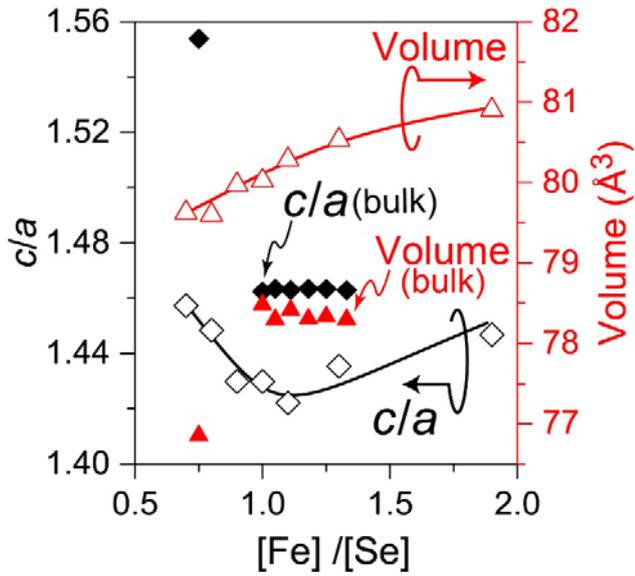

Figure S3. *c*-axis to *a*-axis lattice parameter ratio (*c*/*a*, open black triangles) and cell volume (open red triangles), which were calculated from the values in Figure 4(a), as a function of [Fe]/[Se]. Filled triangles and diamonds indicate the *c*/*a* ratio and volume for some bulk FeSe samples, respectively [30, 31, 33].

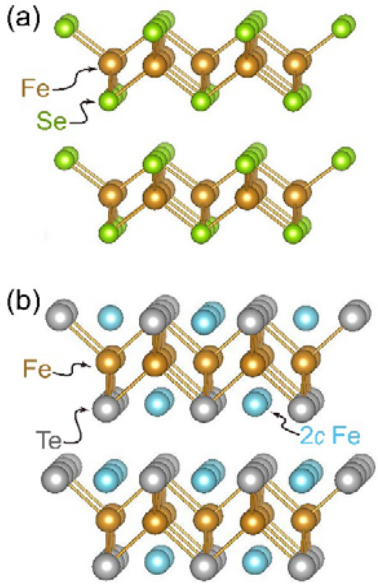

Figure S4. Crystal structure of (a) *β*-FeSe and (b) *β*-Fe$_{1+x}$Te with a space group of *P*4/*nmm* (space group: 129), where brown, green, and gray spheres represent Fe, Se, and Te sites, respectively. In Fe$_{1+x}$Te, excess Fe statistically occupies the interlayer 2*c* sites (blue spheres).

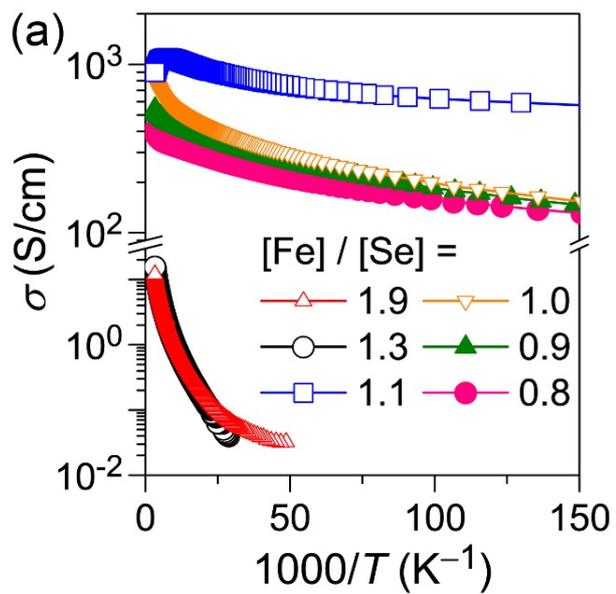
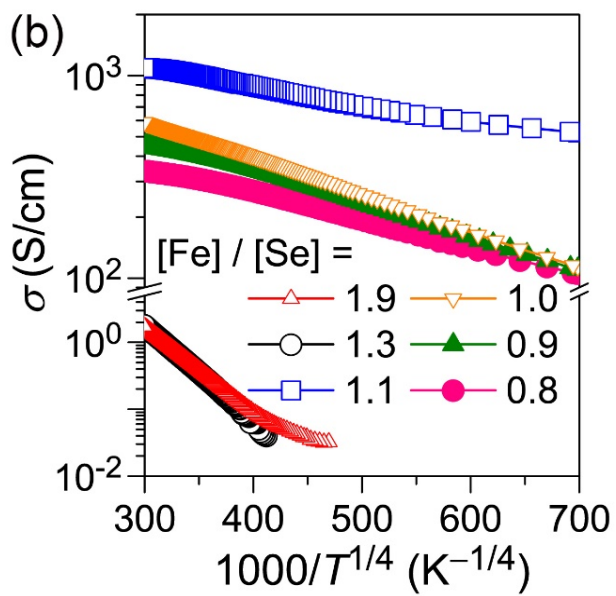

Figure S5. Relationships between $\sigma$ and (a) $T^{-1}$ and (b) $T^{-1/4}$.

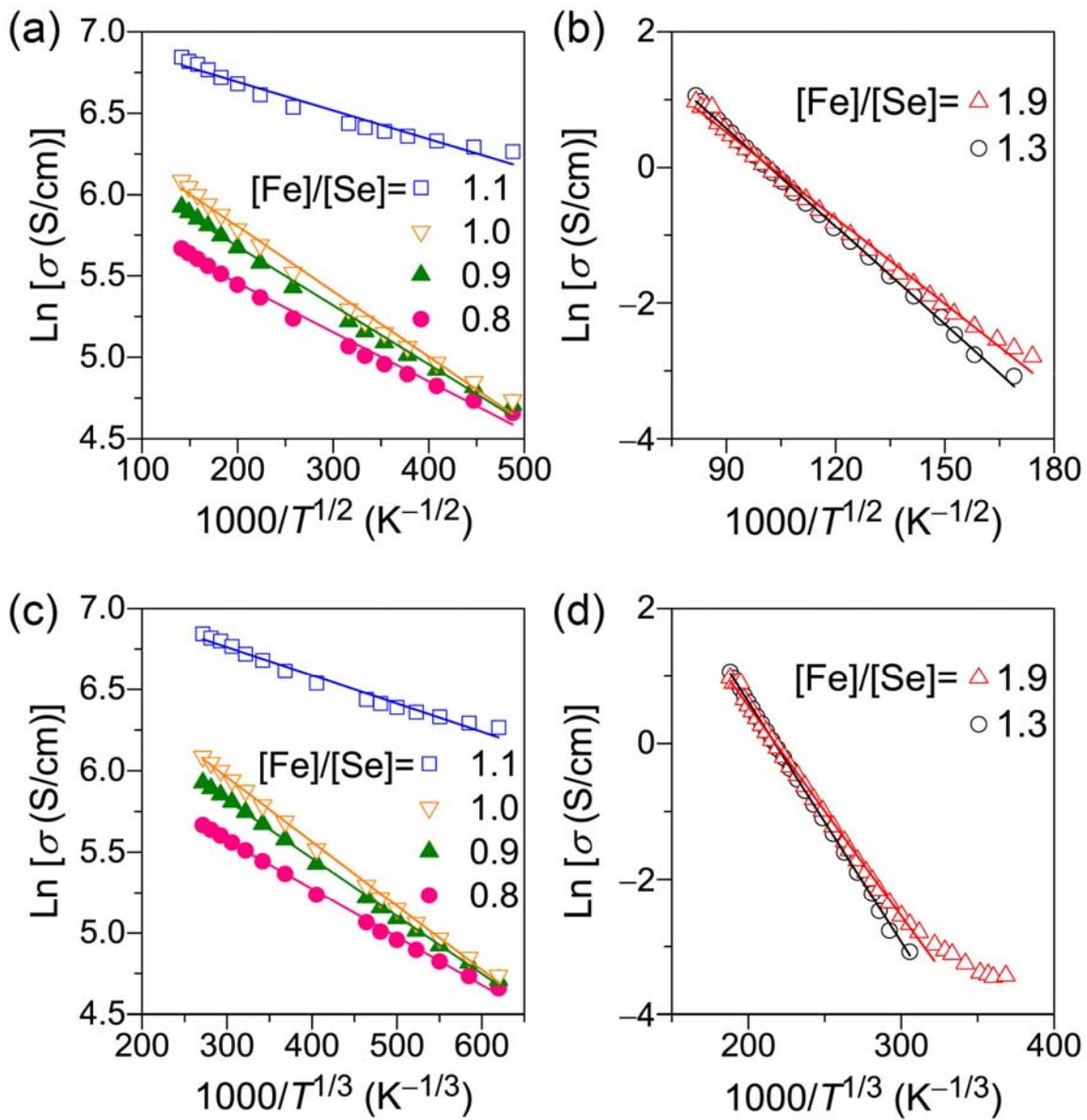

Figure S6. Plots of (a) and (b) Ln$\sigma$ vs. $T^{-1/2}$ and (c) and (d) Ln$\sigma$ vs. $T^{-1/3}$ for FeSe thin films with different compositions.

Table S1. Summary of results obtained from fitting by $T^{-1/n}$ ($n = 2 - 4$). SD is the standard deviation for each fitting.

| [Fe]/[Se] | $\sigma = \sigma_0 exp(-(T_0/T)^{1/2})$ | | | $\sigma = \sigma_0 exp(-(T_0/T)^{1/3})$ | | | $\sigma = \sigma_0 exp(-(T_0/T)^{1/4})$ | | |
|---|---|---|---|---|---|---|---|---|---|
| | $-T_0^{1/2}$ /K$^{1/2}$ | SD /10$^{-5}$K$^{1/2}$ | $T_0$ /K | $-T_0^{1/3}$ /K$^{1/3}$ | SD /10$^{-5}$K$^{1/3}$ | $T_0$ /K | $-T_0^{1/4}$ /K$^{1/4}$ | SD /10$^{-5}$K$^{1/4}$ | $T_0$ /K |
| 0.8 | 3.02 | 7.95 | 9.1 | 2.98 | 3.81 | 26.5 | 3.21 | 2.31 | 106.2 |
| 0.9 | 3.62 | 7.89 | 13.1 | 3.57 | 3.04 | 45.5 | 3.84 | 1.87 | 217.4 |
| 1.0 | 4.01 | 9.86 | 16.1 | 4.01 | 4.31 | 64.4 | 4.26 | 2.26 | 329.3 |
| 1.1 | 1.75 | 9.64 | 3.1 | 1.74 | 7.07 | 5.3 | 1.88 | 6.30 | 12.5 |
| 1.3 | 35.50 | 17.86 | 1260 | 48.12 | 42.10 | 1.1×10$^5$ | 33.12 | 19.08 | 1.2×10$^6$ |
| 1.9 | 30.95 | 34.69 | 958 | 42.19 | 61.46 | 7.5×10$^4$ | 29.47 | 20.34 | 7.5×10$^5$ |

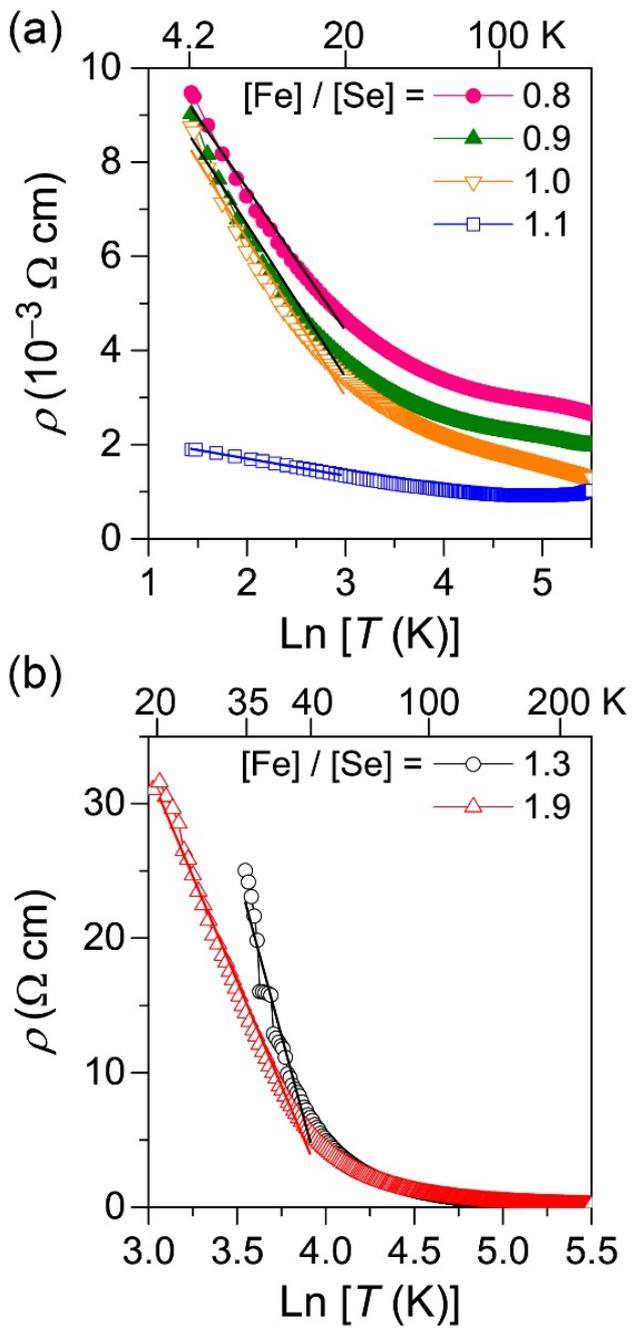

Figure S7. $\rho$ versus ln($T$) for FeSe films with [Fe]/[Se] of (a) 0.8–1.1 and (b) 1.3 and 1.9. Solid lines in each figure are linear least-squares fitting results.